\documentclass{article}
\usepackage[utf8]{inputenc}
\usepackage[english]{babel}
\usepackage{times}
\usepackage{graphicx}
\usepackage{latexsym}
\usepackage{amsmath, amsthm, amssymb, amsfonts}
\usepackage{multirow}
\usepackage{multicol}
\usepackage{fullpage}
\usepackage{bm}
\usepackage{url}

\usepackage[hypcap=false]{caption}
\usepackage{wrapfig}
\usepackage{lineno}

\usepackage{csquotes}

\usepackage[
backend=bibtex,
style=numeric,
sorting=ynt
]{biblatex}
\bibliography{main}

\newcommand{\institute}[1]{\thanks{#1}}

\begin{document}

\title{Chest X-ray lung and heart segmentation based on minimal training sets}
\author{Balázs Maga \institute{Eötvös Loránd University, Pázmány Péter sétány 1/C, Budapest, H-1117 Hungary, email: mbalazs0701@gmail.com}}

\maketitle

\begin{abstract}
As the COVID-19 pandemic aggravated the excessive workload of doctors globally, the demand for computer aided methods in medical imaging analysis increased even further. Such tools can result in more robust diagnostic pipelines which are less prone to human errors. In our paper, we present a deep neural network to which we refer to as Attention BCDU-Net, and apply it to the task of lung and heart segmentation from chest X-ray (CXR) images, a basic but ardous step in the diagnostic pipeline, for instance for the detection of cardiomegaly. We show that the fine-tuned model exceeds previous state-of-the-art results, reaching $98.1\pm 0.1\%$ Dice score and $95.2\pm 0.1\%$ IoU score on the dataset of Japanese Society of Radiological Technology (JSRT). Besides that, we demonstrate the relative simplicity of the task by attaining surprisingly strong results with training sets of size 10 and 20: in terms of Dice score, $97.0\pm 0.8\%$ and $97.3\pm 0.5$, respectively, while in terms of IoU score, $92.2\pm 1.2\%$ and $93.3\pm 0.4\%$, respectively. To achieve these scores, we capitalize on the mixup augmentation technique, which yields a remarkable gain above $4\%$ IoU score in the size 10 setup.
\end{abstract}


\section{Introduction}\label{sec:intro}

All around the world, a plethora of radiographic examinations are performed day to day, producing images using different imaging modalities such as X-ray, computed tomography (CT), diagnostic ultrasound and magnetic resonance imaging (MRI). According to the publicly available, official data of the National Health Service (\cite{england2019diagnostic}), in the period from February 2017 to February 2018, the count of imaging activity was about 41 million only in England. The thorough examination of the vast quantity of these images imposes a huge workload on radiologists, which increases the number of the avoidable human mistakes. Consequently, automated methods aiding the diagnostic processes are sought-after.

The examination of medical images customarily includes various segmentation tasks, in which detecting and pixelwise annotating different tissues and certain anomalies are vital. Common examples include lung nodule segmentation in the diagnosis of lung cancer, lung and heart segmentation in the diagnosis of cardiomegaly, or plaque segmentation in the diagnosis of thrombosis. Even in the case of 2-dimensional modalities, such segmentation tasks can be extremely time-demanding, and the situation gets even worse in three dimension. Taking into consideration that these tasks are easier to formalize as a standard computer vision exercise than the identification of a particular disease, it is not surprising that they sparked much activity in the field of automated medical imaging analysis. 

Semantic segmentation -- that is assigning a pre-defined class to each pixel of an image -- requires a high level of visual understanding, in which state-of-the-art performance is attained by methods utilizing Fully Convolutional Networks (FCN) \cite{long2015fully}. An additional challenge of the field is posed by the strongly limited quantity of training data on which one train machine learning models, as annotating medical imaging requires specialists in contrast to ``real-life'' images. To overcome this difficulty, the so-called U-Net architecture was proposed: its capability to being efficiently trained on small datasets has been demonstrated in \cite{unet}. Over the past few years several modifications and improvements have been proposed on the original architecture, some of which involved different attention mechanisms designed to help the network to detect the important parts of the images. 

In the present paper we introduce a new network primarily based on the ideas of \cite{BCDU-Net} and \cite{oktay2018attention}, to which we refer to as Attention BCDU-Net. We optimize its performance through hyperparameter tests on the depth of the architecture and the loss function, and we demonstrate the superiority of the resulting model compared to the state-of-the-art network presented in \cite{magaal} in the task of lung and heart segmentation on chest X-rays. Besides that, we will also give an insight into two interesting phenomena arising during our research which might be interesting for the AI medical imaging community: one of them is the very small data requirement of this particular task, while the other one is the peculiar evolution of the loss curves over the training.

\section{Deep learning approach}

\subsection{Related work}

As already mentioned in Section \ref{sec:intro}, \cite{unet} introducing U-Nets is of paramount importance in the field. Since then U-Nets have been used to cope with diverse medical segmentation tasks, and numerous papers aimed to design U-Net variants and mechanisms such that the resulting models tackles better the problem considered. Some of these paid primary attention to the structure of the encoder and the decoder -- that is the downsampling and the upsampling path -- of the original architecture. For example in \cite{cole_cnn}, the authors developed a network (CoLe-CNN) with multiple decoder branches and Inception-v4 inspired encoder to achieve state-of-the-art results in 2-dimensional lung nodule segmentation. In \cite{unet++_1} and \cite{unet++_2}, the authors introduced U-Net++, a network equipped with intermediate upsampling paths and additional convolutional layers, leading to essentially an efficient ensemble of U-Nets of varying depths, and demonstrated its superiority compared to the standard U-Net in many image domains. 
Other works put emphasis on the design of skip connections and the way the higher resolution semantic information joins the features coming through the upsampling branch. In \cite{BCDU-Net}, the authors proposed the architecture BCDU-Net, in which instead of the simple concatenation of the corresponding filters, the features of different levels are fused using a bidirectional ConvLSTM layer, which introduces nonlinearity into the model at this point and makes more precise segmentations available. In \cite{oktay2018attention} it has been shown that for medical image analysis tasks the integration of so-called Attention Gates (AGs) improved the accuracy of the segmentation models, while preserving computational efficiency. In \cite{magaal}, this network was enhanced by a critic network in a GAN-like scheme following \cite{wei2018scan}, and achieved state-of-the-art results in the task of lung and heart segmentation. Other attention mechanisms were introduced in \cite{scaunet} and in \cite{scau_net_second}.

\subsection{Our proposal}

The network architecture Attention BCDU-Net we propose is a modification of the Attention U-Net, shown at Figure 1.

\begin{center}
\includegraphics[scale=0.4]{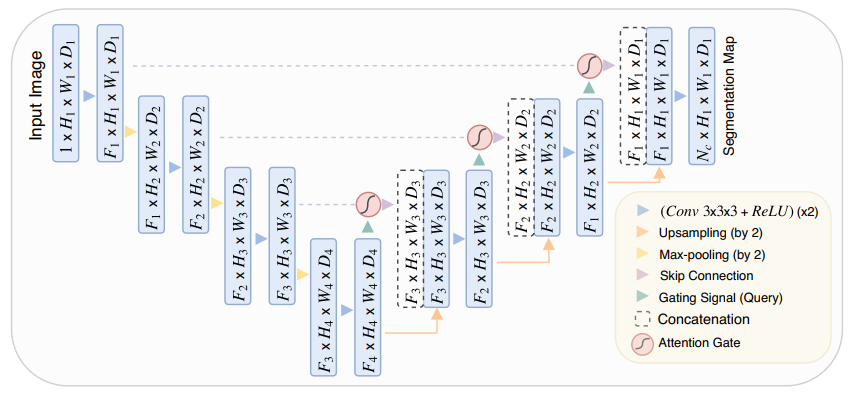}
\captionof{figure}{Schematic architecture of Attention U-Net \cite{oktay2018attention}.}

\medskip

\includegraphics[scale=0.11]{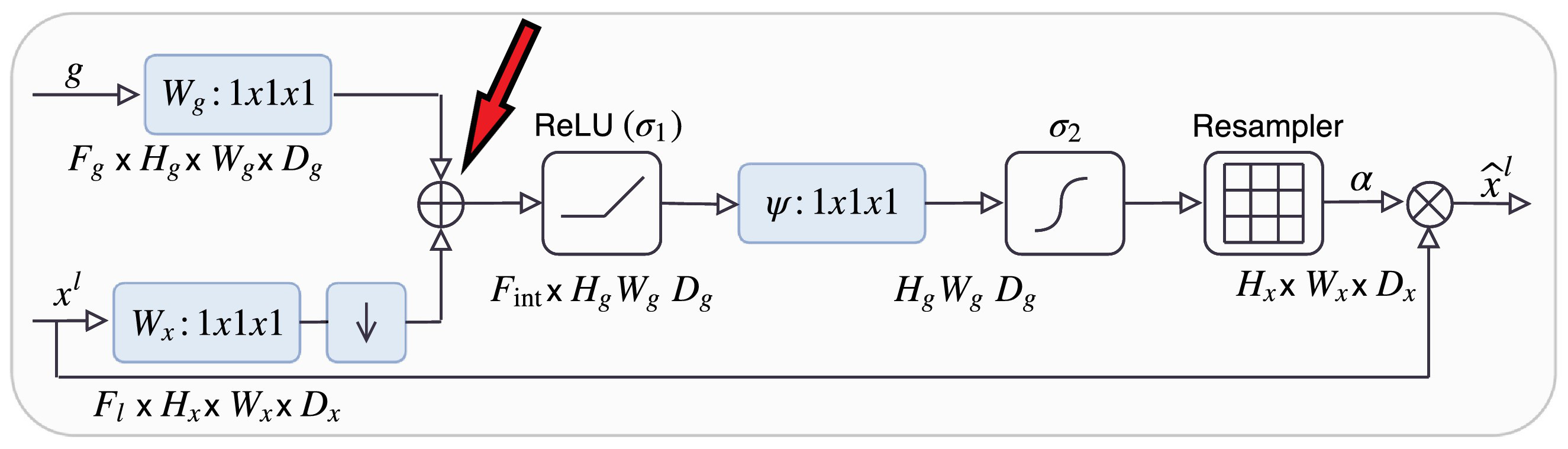}
\captionof{figure}{Schematic figure of the attention gate used in Attention U-Net \cite{oktay2018attention}, the tensor addition to alter is highlighted by an arrow.}
\end{center}

\medskip

In \cite{BCDU-Net}, the authors demonstrated that it is beneficial to use bidirectional ConvLSTM layers to introduce nonlinearity in the step of merging semantic information gained through skip connections and the features arriving through the decoder. This inspired us to modify the attention gates (see Figure 2) in a similar manner, in which these pieces of information are merged via tensor addition, that is a linear operation as well. This addition is replaced by a bidirectional ConvLSTM layer, to which the output of $W_g$ and $W_x$ -- the processed features and the semantic information, respectively -- is fed in this order. We note that to our best knowledge, there is a slight ambiguity about the structure of the resampling steps in the attention gate: while the official implementation is in accordance with the figure, there are widely utilized implementations in which the output of $W_g$ is upsampled instead of downsampling the output of $W_x$ in order to fit their shape. We tested both solutions and did not experience a measurable difference in the performance. We also experimented with the usage of additional spatial and channel attention layers as proposed by \cite{scaunet}, however, we found that it does not improve the performance of our model.

The depth of the network is to be determined by hyperparameter testing. Our tests confirmed that four downsampling steps results in the best performance, however, the differences are minuscule. 

\subsection{Loss function}

A standard score to compare segmentations is the Intersection over Union (IoU): given two sets of pixels $X, Y$, their IoU is
\begin{displaymath}
IoU(X,Y) = \frac{ |X\cap Y|}{|X \cup Y|}.
\end{displaymath}
In the field of medical imaging, Dice Score Coefficient (DSC) is probably the most widespread and simple way to measure the overlap ratio of the masks and the ground truth, and hence to compare and evaluate segmentations. It is a slight modification of IoU: given two sets of pixels $X,Y$, their DSC is
\begin{displaymath}
DSC(X,Y) = \frac{ 2|X\cap Y|}{|X|+|Y|}.
\end{displaymath}
If $Y$ is in fact the result of a test about which pixels are in $X$, we can rewrite it with the usual notation true/false positive (TP/FP), false negative (FN) to be
\begin{displaymath}
DSC(X,Y) = \frac{2TP}{2TP + FN + FP}.
\end{displaymath}
We would like to use this concept in our setup. The class $c$ we would like to segment corresponds to a set, but it is more appropriate to consider its indicator function $g$, that is $g_{i,c}\in \{0,1\}$ equals 1 if and only if the $i$th pixel belongs to the object. On the other hand, our prediction is a probability for each pixel denoted by $p_{i,c}\in[0,1]$. Then the Dice Score of the prediction in the spirit of the above description is defined to be
\begin{displaymath}
DSC = \frac{\sum_{i=1}^N p_{i,c}g_{i,c} + \varepsilon}{\sum_{i=1}^N \left(p_{i,c}+g_{i,c}\right) + \varepsilon},
\end{displaymath}
where $N$ is the total number of pixels, and $\varepsilon$ is introduced for the sake of numerical stability and to avoid divison by 0. The IoU of the prediction can be calculated in a similar manner. The linear Dice Loss (DL) of the multiclass prediction is then
\begin{displaymath}
DL = \sum_{c}\left(1 - DSC_c\right).
\end{displaymath}
A deficiency of Dice Loss is that it penalizes false negative and false positive predictions equally, which results in high precision but low recall. For example practice shows that if the region of interests (ROI) are small, false negative pixels need to have a higher weight than false positive ones. Mathematically this obstacle is easily overcome by introducing weights $\alpha,\beta$ as tuneable parameters, resulting in the definition of Tversky similarity index \cite{tversky1977features}:
\begin{displaymath}
TI_c = \frac{\displaystyle\sum_{i=1}^N p_{i,c}g_{i,c} + \varepsilon}{\displaystyle\sum_{i=1}^N p_{i,c}g_{i,c} + \alpha\displaystyle\sum_{i=1}^N p_{i,\overline{c}}g_{i,c}+\beta\displaystyle\sum_{i=1}^N p_{i,c}g_{i,\overline{c}} + \varepsilon},
\end{displaymath}
where $p_{i,\overline{c}}=1-p_{i,c}$ and $g_{i,\overline{c}}=1-g_{i,c}$, that is the overline simply stands for describing the complement of the class. \\
Tversky Loss is obtained from Tversky index as Dice Loss was obtained from Dice Score Coefficient:
\begin{displaymath}
TL = \sum_{c}\left(1 - TI_c\right).
\end{displaymath}
Another issue with the Dice Loss is that it struggles to segment small ROIs as they do not contribute to the loss significantly. This difficulty was addressed in \cite{abraham2019novel}, where the authors introduced the quantity Focal Tversky Loss in order to improve the performance of their lesion segmentation model:
\begin{displaymath}
FTL = \sum_{c}\left(1 - TI_c\right)^{\gamma^{-1}},
\end{displaymath}
where $\gamma \in [1,3]$. In practice, if a pixel with is misclassified with a high Tversky index, the Focal Tversky Loss is unaffected. However, if the Tversky index is small and the pixel is misclassified, the Focal Tversky Loss will decrease significantly.

In our work we use multiclass DSC and IoU to evaluate segmentation performance. As our initial tests demonstrated that training our network with Focal Tversky loss results in better scores, we will use this loss function. The optimal $\alpha, \beta, \gamma$ parameters should be determined by extensive hyperparameter testing and grid search. We worked below with $\alpha=0.6, \beta=0.4, \frac{1}{\gamma}=0.675$.

\subsection{Dataset and preprocessing}

For training- and validation data, we used the public Japanese Society of Radiological Technology (JSRT) dataset (\cite{shiraishi2000development}), available at \cite{jsrt_dataset}. The JSRT dataset contains a total of 247 images, all of them are in $2048 \times 2048$ resolution, and have 12-bit grayscale levels. Both lung and heart segmentation masks are available for this dataset.

In terms of preprocessing, similarly to \cite{magaal}, the images were resized to the resolution $512 \times 512$ first. As X-rays are grayscale images with typically low contrast, which makes their analysis a difficult task. This obstacle might be overcome by using some sort of histogram equalization technique. The idea of standard histogram equalization is spreading out the the most frequent intensity values to a higher range of the intensity domain by modifying the intensities so that their cumulative distribution function (CDF) on the complete modified image is as close to the CDF of the uniform distribution as possible. Improvements might be made by using adaptive histogram equalization, in which the above method is not utilized globally, but separately on pieces of the image, in order to enhance local contrasts. However, this technique might overamplify noise in near-constant regions, hence our choice was to use Contrast Limited Adaptive Histogram Equalization (CLAHE), which counteracts this effect by clipping the histogram at a predefined value before calculating the CDF, and redistribute this part of the image equally among all the histogram bins. \\

\subsection{Data augmentation}

Concerning data augmentation, we follow \cite{mixup_medical}, in which the method mixup was used to improve glioma segmentation on brain MRI's. This slightly counter-intuitive augmentation technique was introduced by \cite{mixup_original}: training data samples are obtained by taking random convex combinations of original image-mask pairs. That is, for $(x_1, y_1)$ and $(x_2, y_2)$ image-mask pairs, we create a random mixed up pair $x=\lambda x_1 + (1-\lambda)x_2$, $y=\lambda y_1 + (1-\lambda)y_2$, where $\lambda$ is chosen from the beta distribution $B(\delta, \delta)$ for some $\delta\in (0, \infty)$. In each epoch, the original samples are paired randomly, hence during the course of the training, a multitude of training samples are fed to the network. (From the mathematical point of view, as the coefficient $\lambda$ is chosen independently in each case from a continuous probability distribution, the network will encounter pairwise distinct mixed up training samples with probability 1, modulo floating point inaccuracy.) In \cite{mixup_original}, the authors argue that generating training samples via this method encourages the network to behave linearly in-between training examples, which reduces the amount of undesirable oscillations when predicting outside the training examples.

The choice of $\delta$ should be determined by hyperparameter testing for any network and task considered. In \cite{mixup_original}, $\delta\in[0.1, 0.4]$ is proposed, while in \cite{mixup_medical} $\delta=0.4$ is applied. 

\section{Experiments}

\subsection{Training schedule}

In our main tests, the JSRT dataset was randomly split so that 85\% of it was used for training and the rest for validation and testing. This split was carried out independently in each case, enhancing the robustness of our results. Besides that, we also experimented with small dataset training, in which rather modest sets of 10 and 20 X-rays was utilized as training set. (The test set remained the same.) It enabled us to measure the benefits of mixup more transparently. In each of these cases, we trained our network with Adam optimizer: in the former case, for 50 epochs, while in the latter cases for 1000 and 500 epochs, respectively. As these epoch numbers are approximately inversely proportional to the size of the training sets, these choices correspond to each other in terms of training steps.

\subsection{Results}

Table 1 summarizes the numerical results we obtained during the testing of Attention BCDU-Net with different train sizes and choices of $\delta$, while Figure 3-5 display visual results. Note that the highest DSC scores slightly exceed the ones attained by the state-of-the-art, adversarially enhanced Attention U-Net introduced in \cite{magaal} ($97.6\pm 0.5 \%$) and admit higher stability. The effect of augmentation is the most striking in the case of training on an X-ray set of size 10, when the choice $\delta=0.2$ results in a 5\% increase of IoU compared to the no mixup case. In general, we found this case particularly interesting: it was surprising that we could achieve IoU and DSC scores of this magnitude using such a small training set. Nevertheless the predictions have some imperfections, displayed by Figure 3: the contours of the segmentation are less clear and both the heart and the lung segmentation tend to contain small spots far from the ground truth. However, such conspicuous faults are unlikely to occur in the case of the best models for 20 train X-rays (Figure 4), which is still remarkable. The sufficiency of such small training sets is probably due to the relative simplicity of the task. Notably, lung and heart regions admit large similarity across a set of chest X-rays, and they are strongly correlated with simple intensity thresholds. Consequently, even small datasets have high representing potential. We note that as $\delta$ gets smaller, the probability density function of $B(\delta, \delta)$ gets more strongly skewed towards the endpoints of the interval $[0,1]$, which results in mixed up samples being closer to original samples in general. The perceived optimality of $\delta=0.2$ in the small dataset cases show that a considerable augmentation is beneficial and desirable, yet it is unadvised to use too wildly modified samples.

The benevolent effect of mixup gets more obscure as we increase the size of the training set. Notably, the results of different augmentation setups are almost indistinguishable from each other. We interpret this phenomena as another consequence of the similarity of masks from different samples, which inherently drives the network towards simpler representations in the case of a sufficiently broad training set, even without using mixup. 

We also note that in the case of 10 training samples, while the IoU differences between the no mixup and the mixup regime are striking, the gain in DSC is less remarkable. It hints that it is unadvised to rely merely on DSC when evaluating and comparing segmentation models.

\begin{center}
\includegraphics[scale=0.08]{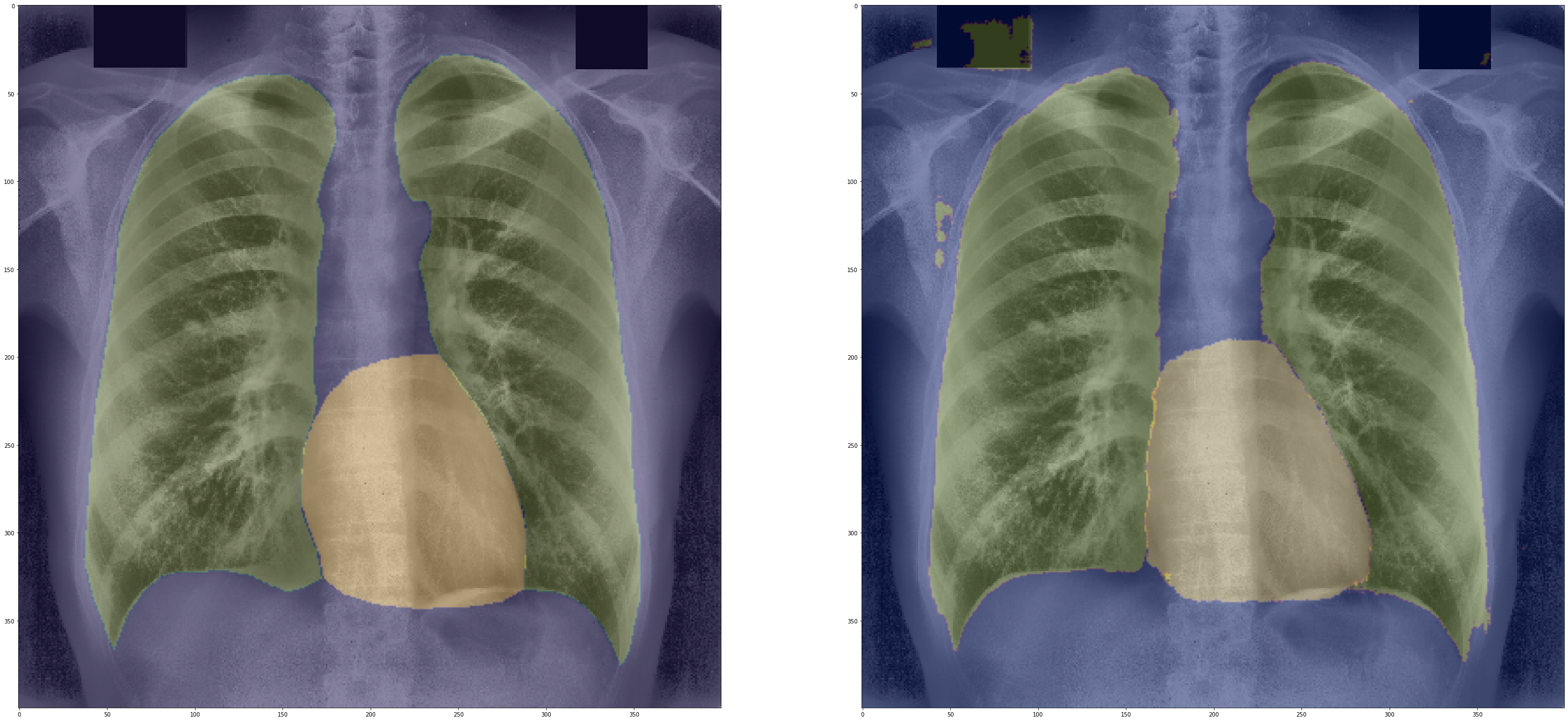}
\captionof{figure}{Ground truth (left) compared to the prediction of the Attention BCDU-Net (right), train size: 10.}

\medskip

\includegraphics[scale=0.08]{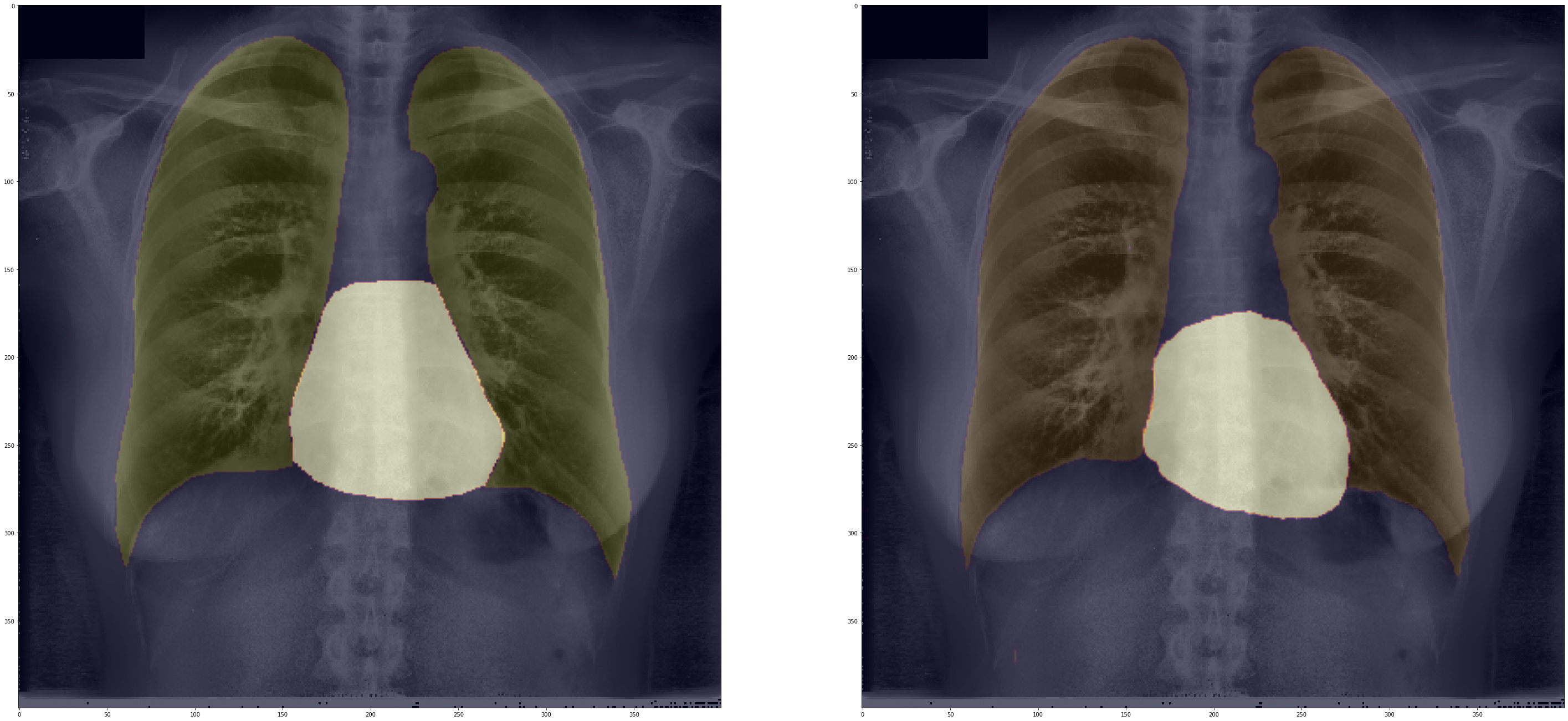}
\captionof{figure}{Ground truth (left) compared to the prediction of the Attention BCDU-Net (right), train size: 20.}

\medskip

\includegraphics[scale=0.08]{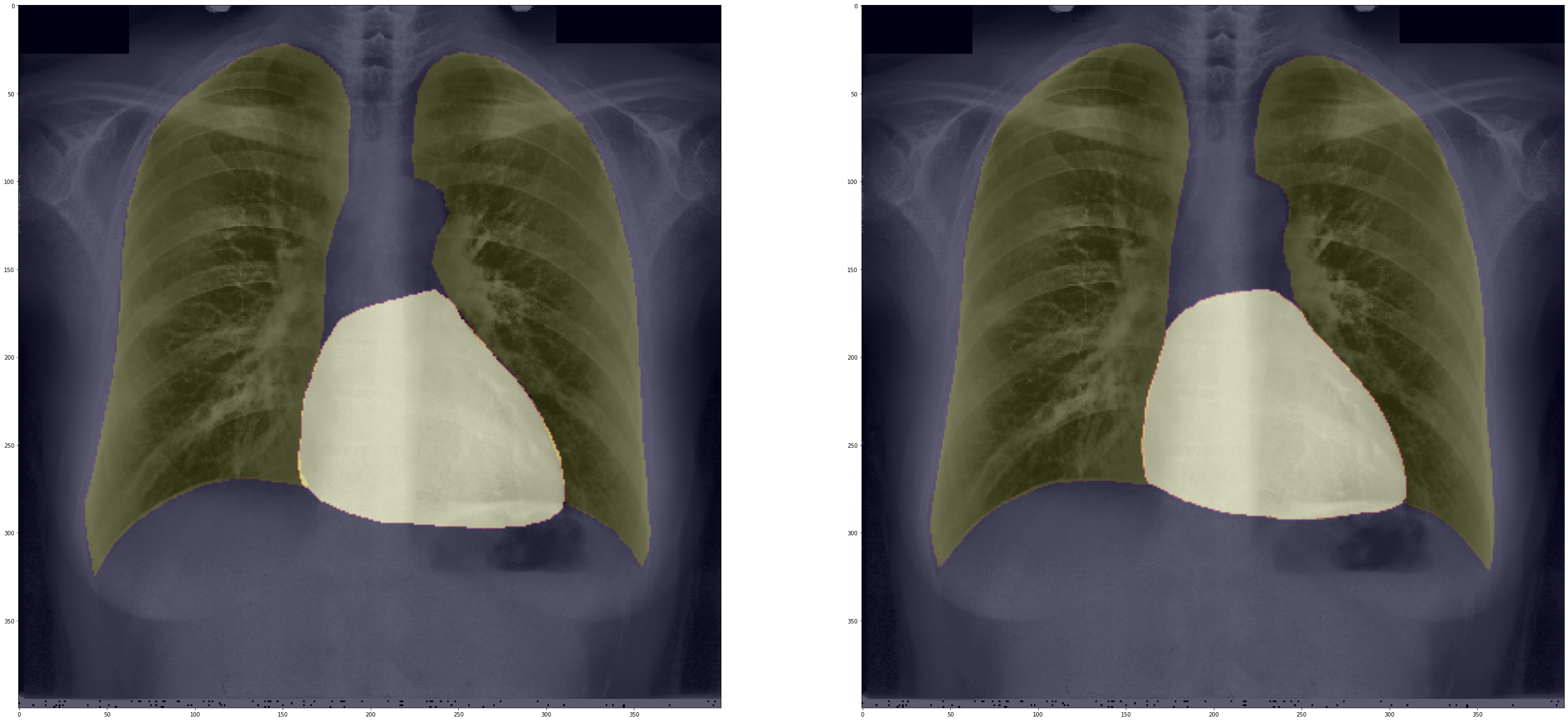}
\captionof{figure}{Ground truth (left) compared to the prediction of the Attention BCDU-Net (right), train size: complete.}

\end{center}


\begin{center}
\begin{tabular}{|c|c|c|c|c|c|c|}
\hline
\rule{0pt}{3pt}
&\multicolumn{2}{|c|}{Train size: 10}&\multicolumn{2}{|c|}{Train set: 20}& \multicolumn{2}{|c|}{Train size: 209 (Complete)}
\\
\hline
\\[-12pt]
 & IoU & DSC & IoU & DSC & IoU & DSC
\\
\hline
\\[-12pt]
No mixup & $87.2 \pm 1.9$\% & $94.8 \pm 1.1$\% & $91.9 \pm 0.6$\% & $96.9 \pm 0.5$\% & $94.9 \pm 0.2$\% & $98.0 \pm 0.1$\%
\\
$\delta=0.1$ & $91.9 \pm 1.3$\% & $96.8 \pm 0.9$\% & $92.5 \pm 0.5$\% & $97.1 \pm 0.5$\% & $95.2 \pm 0.2$\% & $98.1 \pm 0.1$\%
\\
$\delta=0.2$ & $92.2 \pm 1.2$\% & $97.0 \pm 0.8$\% & $93.3 \pm 0.4$\% & $97.3 \pm 0.5$\% & $95.0 \pm 0.1$\% & $98.0 \pm 0.1$\%
\\
$\delta=0.3$ & $91.3 \pm 1.2$\% & $96.5 \pm 1.0$\% & $92.9 \pm 0.5$\% & $97.2 \pm 0.5$\% & $94.9 \pm 0.1$\% & $98.0 \pm 0.1$\%
\\
$\delta=0.4$ & $91.3 \pm 1.4$\% & $96.4 \pm 1.0$\% & $93.0 \pm 0.5$\% & $97.2 \pm 0.4$\% & $94.8 \pm 0.1$\% & $97.9 \pm 0.1$\%
\\
\hline
\end{tabular}
\captionof{table}{Dice scores and IoU scores of Attention BCDU-Net with different mixup parameters}
\end{center}

\begin{figure}[h!]
    \centering
    \includegraphics[scale=0.26]{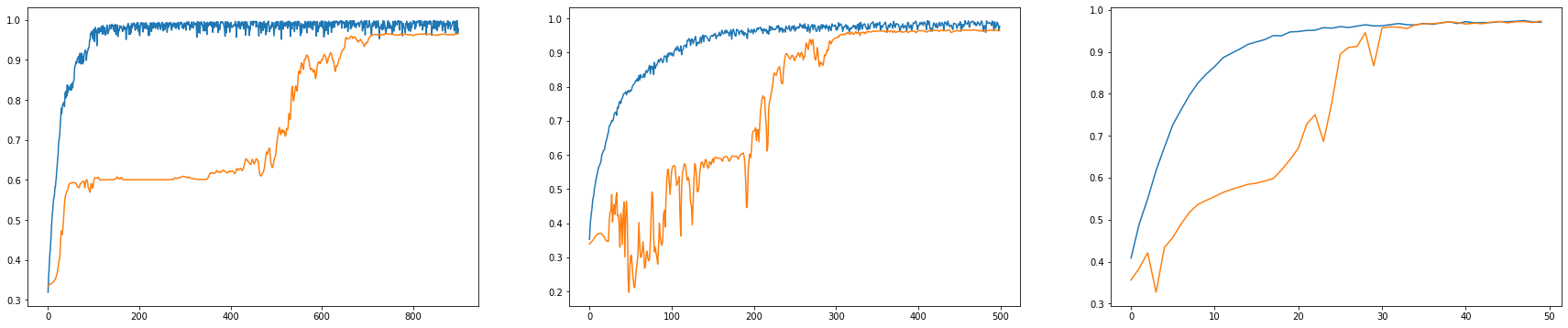}
    \caption{From left to right: the evolution of the train DSC (blue) and the validation DSC (orange) with 10 training samples, 20 training samples, and the complete training dataset, respectively. The IoU curves admit similar patterns.}
    \label{loss_curves}
\end{figure}

We would also like to draw attention to the peculiar loss curves we primarily encountered during the small dataset trainings, as displayed in Figure \ref{loss_curves}. Notably, the curve of the validation DSC flattens far below the also flattening curve of the train DSC, strongly inciting the usage of early stopping. (Train DSC reaches essentially 1 in fact, which is unsurprising with such a small training set.) However, in the later stages the validation DSC catches up, even though the train DSC does not have any room for further improvement. We were especially puzzled by this behaviour in the 10-sized training setup, in which both the train and validation DSC seems completely stabilized after from epoch 50 to epoch 400, yet validation DSC skyrockets in the later stages in a very short amount of time. The same behaviour was experienced during each test run. We have yet to give the intuitive or theoretical explanation for this phenomenon that how the generalizing ability of the model can improve further when it seems to be in a perfect state from the training perspective. We note that these observations naturally led us to experiment with even longer trainings, but to no avail.


\section{Conclusion}

In the present work, we addressed the problem of automated lung and heart segmentation on chest X-rays. We introduced a new model, Attention BCDU-Net, a variant of Attention U-Net equipped with modified attention gates, and surpassed previous state-of-the-art results. We also demonstrated its ability to attain surprisingly reasonable results with strongly limited training sets. Performance in these cases was enhanced using the mixup augmentation technique, resulting in highly notable contribution in the IoU score.

Concerning future work, a natural extension of this work would be adding a structure correcting adversarial network to the training scheme, similarly to \cite{wei2018scan} and \cite{magaal}, and measuring its effect on the performance, especially in the setup of limited training sets. We would also like to give some kind of explanation to the phenomenon of peculiar loss curves. 

\section*{\bf Acknowledgements} The project was supported by the grant EFOP-3.6.3-VEKOP-16-2017-00002.

\printbibliography

\end{document}